# Topography-induced symmetry transition of droplets on quasi-periodically patterned surfaces

Enhui Chen[1,2], Quanzi Yuan[1,2], Ya-Pu Zhao[1,2*]

[1] State Key Laboratory of Nonlinear Mechanics, Institute of Mechanics, Chinese Academy of Sciences, Beijing 100190, China

[2] School of Engineering Science, University of Chinese Academy of Sciences, Beijing 100049, China

[*]Corresponding author: yzhao@imech.ac.cn

**Abstract**

Quasi-periodic structures of quasicrystals yield novel effects in diverse systems. However, there is little investigation on employing quasi-periodic structures in the morphology control. Here, we show the use of quasi-periodic surface structures in controlling the transition of liquid droplets. Although surface structures seem random-like, we find that on these surfaces, droplets spread to well-defined 5-fold symmetric shapes and the symmetry of droplet shapes spontaneously restore during spreading, hitherto unreported in the morphology control of droplets. To obtain physical insights into these symmetry transitions, we conduct energy analysis and perform systematic experiments by varying properties of both liquid droplet and patterned surface. The results show the dominant factors in determining droplet shapes to be surface topography and the self-similarity of the surface structure. Our findings significantly advance the control capability of the droplet morphology. Such a quasi-periodic patterning strategy can offer a new method to achieve complex patterns.

Quasi-periodic structures, being ordered while aperiodic, have brought excitements in diverse fields. For example, these structures have initiated the paradigm shift in crystallography (1), introduced special beauty to architectures (2), and recently extended to soft materials including colloids (3), thin-film systems (4) and self-assembly of molecules (5). These structures were extended to a variety of applications and bring special effects, such as the acoustic and optical wave transport in quasicrystals (6, 7). Here, we employ quasi-periodic structures to morphology control by placing liquid droplets on solid surfaces decorated with these structures.

Wetting and spreading of liquid droplets on solid surfaces were extensively investigated in the past century (8-13) because of their ubiquitous existence in nature and applications. Recently, much research has focused on morphology control of droplets using micro-decorated surfaces (14-21). However, similar to the crystals with periodic structures, which have only 2, 3, 4 and 6-fold symmetries, surfaces patterned with regular micropillars in periodic arrangement can yield droplets with contact areas of corresponding symmetries. More specifically, droplets form contact areas of 2-fold symmetry on surfaces patterned with rectangle array (14, 15) and stretched hexagonal array of cylindrical micropillars (15), of 3-fold symmetry on surfaces with hexagonal array of equilateral triangle micropillars (16), of 4-fold symmetry on surfaces with square array of regular micropillars (14-22), and of 6-fold symmetry on surfaces with hexagonal array of cylindrical micropillars (14, 15). The challenging issue is whether it is possible to produce the wetted area of 5-fold symmetry by orderly arranged pillars. Motivated by the aperiodic but ordered structure of quasicrystals with 5-fold symmetric Bragg diffraction pattern and by such structure presents in nature (Fig. 1a), we use surfaces patterned with two-dimensional quasi-periodic micropillar arrays for droplet spreading. We expect to obtain novel topographic features of droplets on such surfaces in comparison with periodically patterned surfaces.

In this article, we find that droplets spread to 5- and 10-fold symmetric shapes during spreading on the surface structures of a Penrose tiling. Droplets experience remarkable shape transitions of spontaneous loss and restoration of symmetry during spreading, for example, from circularity to 5-fold symmetry, and from 10-fold symmetry to circularity. Furthermore, we provide physical insights into how quasi-periodic structures control the shape formation of a droplet during the spreading process.

To begin with, a droplet of silicon oil, which is non-volatile and can completely wet (23) a quasi-periodic surface, was released at the local symmetrical center of the quasi-periodically patterned surface. The size of the liquid droplet is smaller than the capillary length, $l_{ca} = \sqrt{\gamma/\rho g}$ ($\gamma$, $\rho$ and $g$ are the liquid surface tension, density, and gravitational acceleration, respectively), such that the gravity is negligible. Figure 1b shows the surface, on which micropillars are arranged in the pattern of a Penrose tiling (24). This pattern has 5-fold symmetry and is self-similar (the same patterns appear at different scales) with many local symmetrical centers (Fig. 1c-d). The pillar density and roughness are uniform in a long-range area of the pattern.

The temporal evolution of the wetted area of a droplet of silicon oil placed on the quasi-periodic surface with the pattern of a Penrose tiling is shown in Fig. 2. These results demonstrate for the first time that the shape of the wetted area of a droplet on such a surface went through a circle, pentagon, decagon, etc. The quasi-periodic surface with the pattern of a Penrose tiling has made it possible to produce the wetted area with well-defined and reproducible 5-fold and 10-fold symmetries. There exist shape transitions of spontaneous symmetry loss and restoration during the spreading of the droplet from circularity to 5-/10-fold symmetry, from 5-/10-fold symmetry to 10-/5-fold symmetry, and from 5-/10-fold symmetry to circularity. Note that a circular wetted area can only transform into an area with a specific fold symmetry on a periodically patterned surface.

In general, the change of the symmetric characteristic of the wetting area during the spreading of a droplet on a surface patterned with pillars is dependent on the local interaction between the liquid-solid contact line and the pillars. We followed the motion of local contact line through two adjacent pillars, as shown in Fig. 3a, at different states of the contact line. For the spreading of the droplet with the capillary number (Ca=$v\eta/\gamma$~$10^{-1}$), the Reynolds number (Re=$\rho vl/\eta$~$10^{-4}$), the Weber number (We=$\rho v^2 l/\gamma$~$10^{-6}$) and the Bond number (Bo=$\rho g l^2/\gamma$~0.07) much less than 1, it is the surface tension and the interaction between the contact line and pillars that drive the contact line. Here, $v$, $\eta$ and $l$ are the velocity of the contact line, the dynamic viscosity of liquid and the characteristic length scale (e.g. the radius of the initial droplet or the pillars), respectively.

After placing a droplet on a quasi-periodically patterned surface, the droplet driven by the difference of surface tensions starts to spread in radial direction. If the interaction between the contact line and the pillars is negligible, the segment of the contact line of a circular shape experiences the same driving force. All the segment of the contact line moves at the same speed, and the contact line remains circular, as shown in Fig. 3a-i, if there is no local disturbance to the motion. With the contact line approaching the nearby pillars, the interaction between the contact line and the pillars causes the jump-in contact of the contact line to the pillars (pinning (10)). The good wettability between the liquid and the pillars allows the contact line to pass through the pillars at a higher speed due to the Concus-Finn effect (wetting liquids accelerate in a wedge) (25, 26). This process is validated by experiments using a high-speed camera combined with a high-magnification microscope (Fig. 3b). This trend leads to the change of local curvature from concave down to concave up, as shown in Fig. 3a-ii, and the change of the local resultant surface tension. This change increases the driving force for the motion of the concave contact line, and accelerates the motion of the segments of the contact line with concave up to catch up the segments with concave down. Then the contact line restores to circular shape (de-pinning (10)), as shown in Fig. 3a-iii. The transition from pinning to de-pinning during spreading makes it possible for the contact line to deform and restore in the sequences of circle, pentagon, decagon, pentagon, circle, and decagon, as shown in Fig. 2i-viii. This result suggests that one can manipulate the temporal evolution of the wetting area by using patterned surfaces and controlling the transition between pinning and de-pinning of the contact line.

From Fig. 2 ix-xiii, it is interesting to note that the droplet spreads to well-defined pentagonal and decagonal shapes over a large spreading area in a region with random-like distribution of the pillars around the contact line. In contrast to the edges of the wetting area in the early stage of the spreading (i to viii), the edges of the wetting area in the late stage of spreading (ix to xiii) seem wider. Such behavior likely reveals the effect of the radius associated with the pressure inside the droplet (27). For a droplet in a large wetted area ($R \gg R_0$, $R$ being the radius of wetted area and $R_0$ being the initial radius of the droplet), the contribution of the pressure to the motion of the contact line is negligible, and the motion of the contact line is solely controlled by surface tensions and the pillars.

To understand the symmetry transition in the late stage of spreading, we investigated the surface energy of the liquid-solid system consisting of a droplet and a patterned surface. The change of the total surface energy associated with wetting is

$$E = \gamma A - (S_i + \gamma)(A_p + n \cdot 2\pi rh) = \gamma(A - A_p - n \cdot 2\pi rh) - S_i(A_p + n \cdot 2\pi rh).$$

Here, $n$ is the number of wetted pillars, $A$ and $A_p$ are the liquid-vapor surface area and corresponding projected area (wetted area), respectively, $r$ and $h$ are the radius and height of pillars, respectively, and $S_i = \gamma_{sv} - (\gamma_{sl} + \gamma)$ is spreading coefficient (23). As suggested by de Gennes (23) in the energy minimization principle, the last term $S_i(A_p + n \cdot 2\pi rh)$ does not contribute to the macroscopic spreading (12). The parameters controlling the shape transition of the droplet are the surface tension of $\gamma$ and topographic structures. In addition, the velocity of the contact line is dependent on the liquid viscosity (28, 29). Thus, the shape transition of the droplet is also dependent on the liquid viscosity of $\eta$. Though a more accurate model is required in the energy analysis, the elements relevant to the symmetry transition are not altered.

To determine the dominant factors controlling the formation of symmetrical shapes of droplets during wetting, a series of experiments were performed using liquids of different surface tensions, viscosities, and surfaces with different topographical parameters. The pillar densities (15) of $2r/a$ and $2r/a'$ were in a range of 0.03-0.20, the aspect ratios (14) of $h/(a-2r)$ and $h/(a'-2r)$ were in a range of 0.03-0.10, the capillary number of $v\eta/\gamma$ was in a range of $10^{-3}$-$10^{-1}$, and surfaces were treated to achieve complete wetting. Figure 3c summarizes the symmetrical shapes of the droplets formed during spreading. It is evident that the symmetrical transition of the wetted area is controlled by the dimensionless length of $R/a$ with $R$ being the distance from the center of the wetting area to the contact line along one of the principle axes of symmetry. For those with stack of symmetrical shapes, it is the surface patterns that control the transition of the wetted area instead of the liquid viscosity, surface tension and surface topological parameters. Comparing with the wetted areas of well-defined droplet shapes to self-similarity of surface patterns, one can note that the edges of the pentagonal and decagonal droplets exactly overlap with the edges of the pentagonal and decagonal surface patterns of large scales, respectively, as shown in Fig. 3d. Future investigations can likely expand the ideas to partial wetting and/or gravity-dominated spreading systems.

We further extended similar approach to quasi-periodically patterned surfaces of 6- and 8-fold symmetries, respectively (for the patterns, see Refs. (30, 31) and Fig. 4a-b,d-e). Figure 4c,f shows temporal evolution of the wetted area after placing droplets at a local symmetrical center. There exist spontaneous transitions of the wetted areas during the spreading of the droplets from one symmetrical shape to another symmetrical shape. Using the quasi-periodic pattern of 6-fold can transfer a circular droplet to a droplet of 6-fold symmetry, to a circular droplet, to a droplet of 12-fold symmetry, and to a droplet of 6-fold symmetry. Using the quasi-periodic pattern of 8-fold, a circular droplet can transfer to a droplet of 8-fold symmetry, to a circular droplet, and to a droplet of 8-fold symmetry. In both cases, the edges of the second and last droplet shapes overlap with the edges of the corresponding self-similar surface patterns, respectively.

To conclude, the consequences of quasi-periodic surface patterns to liquid droplets placed at local symmetric centers of patterned surfaces are: the symmetry transition of the shapes of wetted area (1) follows the rotational symmetry of surface patterns; (2) transforms spontaneously with symmetry loss and restoration during the droplet spreading; (3) is dependent on the topographic features and self-similarity of surface patterns.

In particular, producing droplets with the wetted area of 5-fold symmetry has considerably advanced the control ability of the morphologies of droplets reported in the literature. The strategy developed in this work can greatly improve the design of devices such as microfluidic chips (32) and topological insulators for sound and light (33). The spontaneous symmetry transition of the wetting area during the spreading of droplets from one pattern to another pattern can offer insights into a variety of applications of low cost fabrication, including ink-jet printing (34) in 3D and 4D printing and conductive patterns in electronic devices (35). Our results may open the way of using aperiodic surface structure to generate desired liquid morphology.

**Methods**

Micro-patterned surfaces with quasi-periodic pillar arrays were first prepared. Surface treatments of the prepared surfaces were then performed to make liquids completely wet the surfaces. A liquid droplet was placed at one of the local symmetrical centers of the treated, patterned surfaces under an optical microscope. A digital camera was used to monitor the spreading process. The following provides the detailed steps.

**Surface preparation.** Quasi-periodic patterns, which were designed by CAD, were transferred to photomasks and next to substrates using photolithography. Both glass and silicon wafers were used as substrates to demonstrate that the observed behavior is independent of substrates. For glass wafers, micro pillars of SU-8 (Photoresist SU-8 2010, MicroChem) were fabricated using standard contact photolithography (6/350/NUV/DCCD/BSV/M, ABM). For silicon wafers, micro pillars were fabricated with standard contact photolithography followed by deep reactive ion etching (Suzhou In Situ Technology Company, Suzhou). The patterned Si-wafers were ultrasonically cleaned in acetone, alcohol and deionized water, successively, and followed with blow drying with $N_2$. Surface-plasma treatments of the wafers were conducted in air for 120 s (Plasma Cleaner PDC-002, Harrick Plasma). Glycerol droplets were placed on the treated surface of the patterned wafers to achieve complete wetting.

**Materials.** Various patterns and liquids were used to determine the control factors. For surface patterns, the geometrical parameters are: $r$ = 7.5-25 μm, $h$ = 10 μm, and $a$ = 200-400 μm. For liquid droplets, surface tension is in a range of 20.1 to 63.3 mN m$^{-1}$, and viscosity is in a range of 10 to 1412 mPa s. Table 1 lists the properties of the liquids we used at 20 °C.

Table 1. Properties of liquids at 20 °C

| Liquid | Viscosity (mPa s) | Surface tension (mN m$^{-1}$) | Density (Kg m$^{-3}$) |
|---|---|---|---|
| Silicon oil (KF-96, Shin-Etsu)(36) | 10 | 20.1 | 935 |
| | 52 | 20.8 | 960 |
| | 515 | 21.1 | 970 |
| | 1030 | 21.1 | 970 |
| Glycerol | 1412 | 63.3 | 1261 |

**Experiments.** Spreading experiments were performed at a temperature of 19.3-23.9°C and a relative humidity of 20-24%. Liquid droplets were placed at local symmetrical centers of the patterned surfaces under an optical microscope, and the spreading process was recorded by a digital camera. All droplets completely wet the surfaces due to the low surface tension of silicon oil and freshly plasma-treated surfaces (within 15 min) for glycerol. For SU8-Glass wafers, experiments were conducted using an inverted microscope (IX71, Olympus) to reduce the shadow effect of micropipette. For silicon wafers, experiments were performed with a digital microscope (KH-8700, Hirox) using coaxial light to identify the moving contact line clearly. For the process of the jump-in contact of a contact line to pillars, a high-speed camera (Fastcam SA8, Photron) combined with a microscope (KH-8700, Hirox) were used.

34. Su M, Li F, Chen S, Huang Z, Qin M, Li W, Zhang X, Song Y (2016) Nanoparticle based curve arrays for multirecognition flexible electronics. *Adv. Mater.* 28(7):1369-1374.
35. Zhang Z, Zhang X, Xin Z, Deng M, Wen Y, Song Y (2013) Controlled inkjetting of a conductive pattern of silver nanoparticles based on the coffee-ring effect. *Adv. Mater.* 25(46):6714-6718.
36. Shin-Etsu Chemical Company, http://www.silicone.jp/catalog/pdf/kf96_j.pdf.

**Acknowledgements** This work was jointly supported by the National Natural Science Foundation of China (NSFC, Grant Nos. U1562105, 11372313 and 11672300), and the CAS Strategic Priority Research Program (Grant No. XDB22040401).

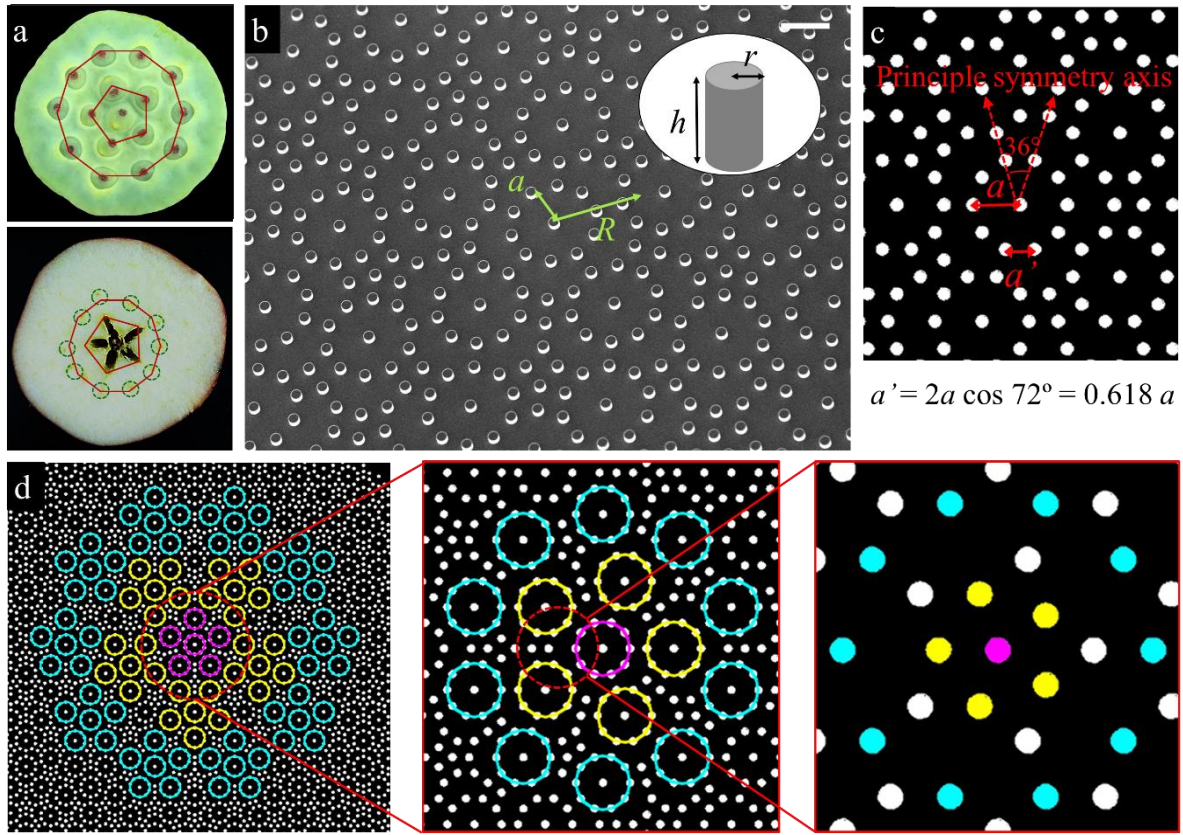

Fig. 1. Quasi-periodic structures. (*a*) Part of quasi-periodic structure in nature. Up: seedpod of a lotus. Down: apple transection with one central hole, five median carpellary bundles, ten petal bundles (labelled in blue dotted circles) and pentagonal outer contour. (*b*) Scanning electron microscope image of a quasi-periodic surface, on which pillars are arranged at intersections of Penrose tiling. The geometry of patterned surfaces is characterized by [*r h a*], where *r* and *h* are pillar radius and height, respectively, *a* denotes as the distance from local symmetry center to the adjacent ten pillars. Inset: schematic of the pillar geometry. Droplet wetted area is characterized by droplet spreading distance (*R*) along one principle symmetry axis. Scale bar, 200 μm. (*c*) Illustration of the topography of the 2D surface pattern. *a* and *a'* are the nearest neighbor distances between pillars. (*d*) Illustrations generated by CAD software showing self-similarity of the five-fold quasi-periodic surface pattern. The similar elements (one-five-ten units successively from center to outer) appear in different scales as labelled in colorful circles or dots.

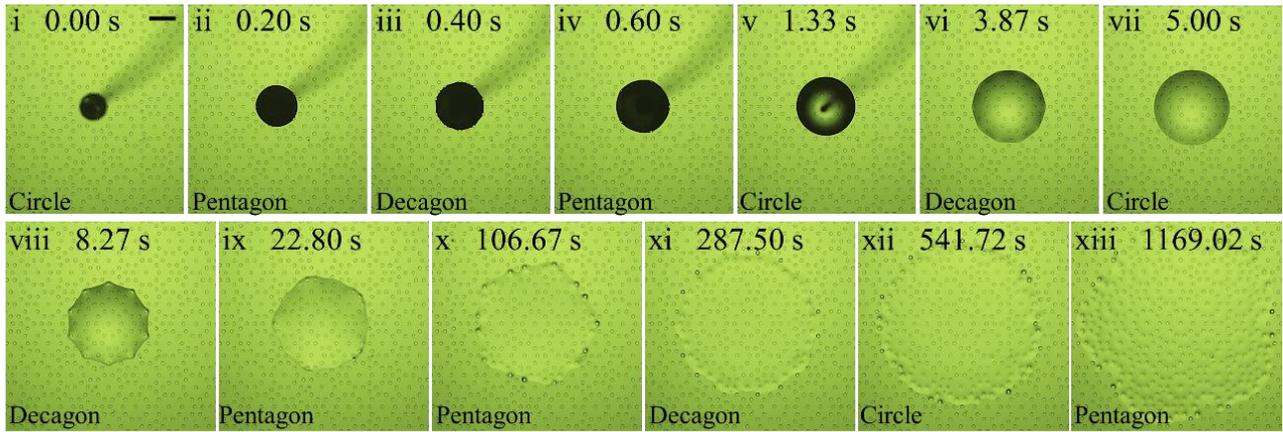

Fig. 2. The temporal evolution of a spreading droplet. A silicon oil droplet (viscosity $\eta$=1030 mPa s) spreading on a quasi-periodic surface with geometry parameters [$r$ $h$ $a$]=[25 10 200] μm. The initial spherical droplet spreads to 5-fold, 10-fold, 5-fold, circular, 10-fold, 5-fold, 10-fold and 5-fold symmetric wetted area successively. Scale bar, 400 μm.

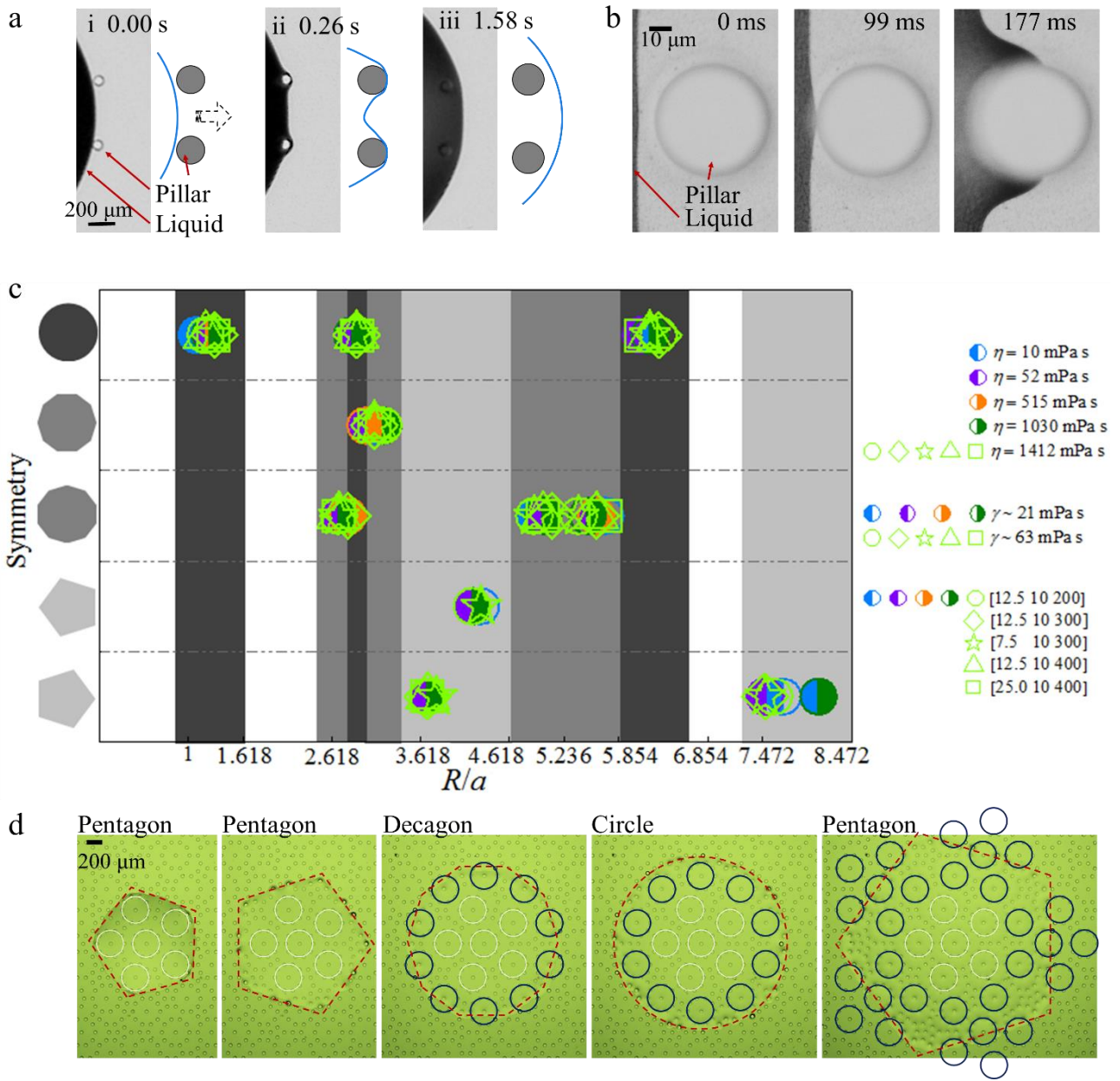

Fig. 3. The behavior and position of contact line control shape transition. (*a*) The behavior of the contact line when passing through two adjacent micro pillars. (*b*) Jump-in contact of a contact line to a pillar. A circular contact line (0 ms) jumps into contact with a pillar when approaching a pillar (99 ms). The contact line then passes through the pillar at a larger speed than that without the pillar (177 ms). (*c*) Schematic of symmetry transition and wetted area of a droplet. Each type of symbol represents an experiment of a liquid-patterned surface pair. Each column in grey indicates the corresponding symmetry of a droplet during spreading. Geometry parameters of surfaces are [*r h a*]

μm. (*d*) Illustration of droplet shapes in the late stage of spreading with respect to surface topography of self-similarity. The edges of the pentagonal and decagonal droplets exactly overlap with the edges of the pentagonal and decagonal surface patterns of large scales, respectively.

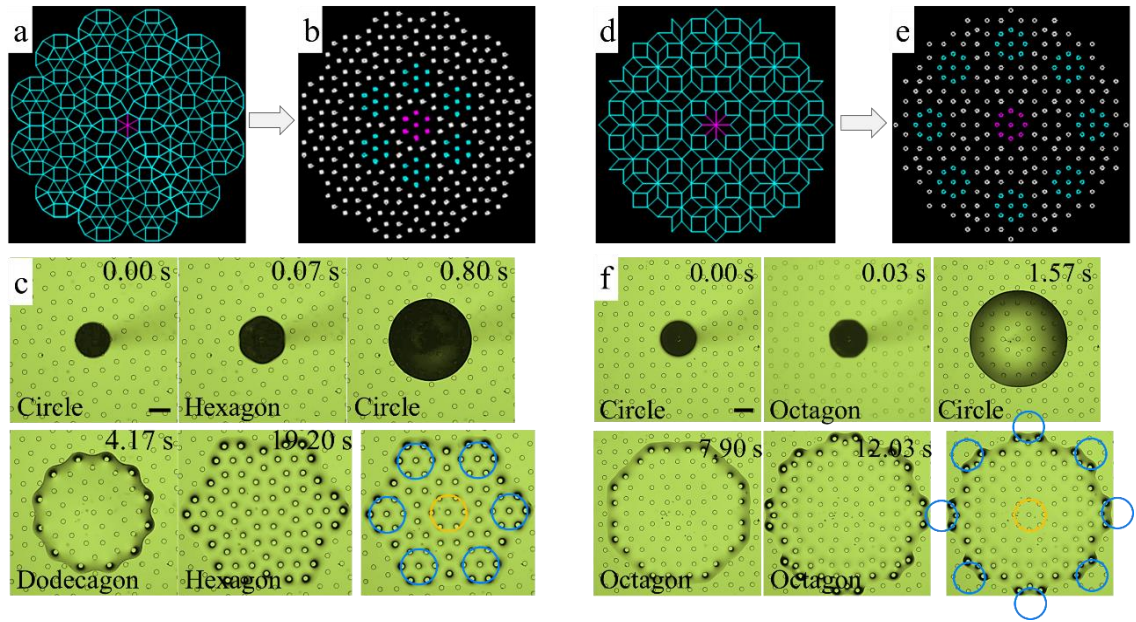

Fig. 4. Quasi-periodic surface patterns and symmetry transitions of spreading droplets. (*a-c*): (*a*) Stampfli tiling (30) generated by CAD software. (*b*) Illustration of quasi-periodic patterns of 6-fold symmetry, on which pillars are arranged at intersections of the Stampfli tiling. (*c*) The temporal evolution of a spreading droplet on a quasi-periodically patterned surface of 6-fold symmetry. Geometry parameters of the surface are [*r h a*]=[25 20 200] μm. (*d-f*): (*d*) Ammann-Beenker tiling (31) generated by CAD software. (*e*) Illustration of quasi-periodic patterns of 8-fold symmetry, on which pillars are arranged at intersections of the Ammann-Beenker tiling. (*f*) The temporal evolution of a spreading droplet on a quasi-periodically patterned surface of 6-fold symmetry. Geometry parameters of the surface are [*r h a*]=[25 20 200] μm. Self-similarity of patterns is indicated by colorful dots in (*b*, *e*). Silicon oil droplets of viscosity $\eta$=1030 mPa s were used. Scale bar, 200 μm.